# Dynamics of coupled vortices in layered magnetic nanodots

K. Yu. Guslienko,[*] K. S. Buchanan, S. D. Bader, and V. Novosad

*Materials Science Division and Center for Nanoscale Materials, Argonne National Laboratory,*

*Argonne, Illinois 60439*

## Abstract

The spin dynamics are calculated for a model system consisting of magnetically soft, layered nanomagnets, in which two ferromagnetic (*F*) cylindrical dots, each with a magnetic vortex ground state, are separated by a non-magnetic spacer (*N*). This permits a study of the effects of interlayer magnetostatic interactions on the vortex dynamics. The system was explored by applying the equations of motion for the vortex core positions. The restoring force was calculated taking into account the magnetostatic interactions assuming a realistic "surface charge free" spin distribution. For tri-layer *F/N/F* dots with opposite chiralities and the same core polarizations (lowest energy state), two eigenmodes are predicted analytically and confirmed via micromagnetic simulations. One mode is in the sub-GHz range for submicron dot diameters and corresponds to quasi-circular rotation of the cores about the dot center. A second mode is in the MHz range corresponding to a small amplitude rotation of the mean core position. The eigenfrequencies depend strongly on the geometrical parameters of the system, suggesting that magnetostatic effects play a dominant role in determining the vortex dynamics.

---

[*] Corresponding author. E-mail: gusliyenko@anl.gov

Magnetically soft submicron structures have received considerable attention due to their tendency to support a stable vortex magnetization state. It is well established that a vortex (soliton-like) state appears for large enough lateral particle dimensions.[1-3] The vortex consists of an in-plane, flux-closure magnetization distribution and a central core whose magnetization is perpendicular to the dot plane. The vortex magnetization distribution leads to considerable modification of the spin excitation spectrum in comparison with a uniformly magnetized dot. In particular, a low-frequency mode of displacement of the vortex as a whole appears.[4, 5] Magnetostatic fields play an important role in determining the dynamic vortex excitations in magnetic dots.

While the conditions necessary for a single-layer disk to support a vortex are well known, as is the reversal through vortex nucleation, displacement and annihilation,[2,3,6] the properties of stacked ferromagnetic (*F*) disks separated by a thin, non-magnetic (*N*) spacer (*F/N/F* structures) have received less attention. Micromagnetic simulations indicate that the disks will each support oppositely directed vortices at remanence.[7] This double vortex state can be created by choice of disk parameters or by including a small antiferromagnetic interlayer exchange contribution.[7] The dynamic properties of *F/N/F* disks have not yet been considered. The disks are couple through both magnetostatic interactions and interlayer exchange (depending on the spacer material and thickness), which will result in unique vortex oscillations.

It is only recently that the vortex excitation modes of single layer F structures have been observed experimentally, i.e. using Brillouin light scattering (BLS),[8] time-resolved Kerr effect[9,10] and X-ray magnetic circular dichroism techniques.[11,12] The theory of vortex excitations in small cylindrical particles has been developed in Ref. 5, 8, 13 and micromagnetic simulations have been reported in Ref. 13-15. These works explore the lowest order vortex translation mode as well as other modes with radial and azimuthal symmetry. In previous studies on vortex dynamics, the only interaction effects that have been considered are weak lateral interactions between dots arranged on a two-dimensional (2D) plane.[16] The approach[16] is based on the simplified "rigid vortex" model, which assumes that the vortices are rigid and do not deform as they move. In reality, however, when the vortex moves the spin configuration changes to reduce stray fields. This discrepancy leads to overestimation of the eigenfrequencies and the effects of inter-dot interactions on the vortex core trajectories have not been considered.[16]

In this paper, analytical and numerical investigations of the vortex oscillation modes are presented for magnetically soft tri-layer *F/N/F* disks consisting of two identical cylindrical magnetic dots with radii *R* and thicknesses *L* separated by spacer *N* of thickness *d* (Fig. 1). The effects of strong coupling between vertically stacked disks on the vortex dynamic behavior are explored assuming a realistic "surface charge free" spin distribution, and the vortex trajectories for the ground state dynamics are explicitly tracked. The magnetization distribution in each *F* layer is considered to be 2D. We limit our analysis to the case of magnetic vortices with identical core polarizations and opposite chiralities, assuming that the each layer supports the topologically simplest vortex state with a centered core at remanence. This state has the lowest energy among all possible states for sub-micron tri-layer dots with a thin spacer[7] and should also be the simplest state to observe experimentally.

The theoretical description of the vortex oscillations utilizes the effective equation for vortex collective coordinates.[17] We focus on the vortex translation modes, where the vortex center (a topological soliton) oscillates about its equilibrium position in each layer. These modes have the lowest frequencies in the vortex excitation spectra. The 2D (independent of the *z*-coordinate along the dot thickness) magnetization distributions $\mathbf{m}_j(\boldsymbol{\rho},t) = \mathbf{M}_j(\boldsymbol{\rho},t)/M_{sj}$, $\mathbf{m}_j^2 = 1$ are assumed for each of the *F* dots (*j*= 1, 2 is the *F* layer number, $M_{sj}$ is the saturation magnetization) and the angular parameterization for the dot magnetization components $m_{zj} = \cos\Theta_j$, $m_{xj} + im_{yj} = \sin\Theta_j \exp(i\Phi_j)$, $\Phi_j = q_j\varphi + \Phi_{0j}$ is used. The phase is $\Phi_{0j} = C_j\pi/2$, where $C_j = \pm 1$ describes the vortex chirality. The signs $C_j = \pm 1$ correspond to counter-clockwise (clockwise) rotation of the vectors **m**$_j$ in the dot plane. The total energy of a vortex does not depend on chirality for an isolated disk; however, the energy for coupled disks depends on the chirality product $C_1C_2$. The integer parameter $q_j$ is the vortex topological charge (vorticity),[17,18] and $p_j = \pm 1$ is the vortex core polarization (direction of the $m_z$ component in the vortex center $\rho = 0$). In some sense the solitons can be represented as quasi-particles and can be characterized by their coordinates $(\mathbf{X}_1, \mathbf{X}_2)$, masses and other collective variables.

In order to describe the translation modes of the vortex motion we apply Thiele's equations[4] for each layer, neglecting the small vortex masses:

$$\mathbf{G}_1 \times \frac{d\mathbf{X}_1}{dt} - \frac{\partial W(\mathbf{X}_1, \mathbf{X}_2)}{\partial \mathbf{X}_1} = 0, \quad \mathbf{G}_2 \times \frac{d\mathbf{X}_2}{dt} - \frac{\partial W(\mathbf{X}_1, \mathbf{X}_2)}{\partial \mathbf{X}_2} = 0, \quad (1)$$

where $\mathbf{X}_j = (X_j, Y_j)$ is the vortex center position in j-th layer, $W(\mathbf{X}_1, \mathbf{X}_2) = W_1(\mathbf{X}_1) + W(\mathbf{X}_2) + W_{int}(\mathbf{X}_1, \mathbf{X}_2)$ is the potential energy of the vortices shifted from their equilibrium positions $\mathbf{X}_j = 0$. Here the interaction energy term $W_{int}(\mathbf{X}_1, \mathbf{X}_2)$ includes only magnetostatic contributions. Inclusion of interlayer exchange effects could be achieved by rescaling some of the magnetostatic coupling energy terms but would not affect the form of the result.

The first term in Eq. (1) is the gyroforce determined by the vortex magnetization distribution in equilibrium.[18] The gyroforce is proportional to the gyrovector $\mathbf{G}_j = -G_j \hat{\mathbf{z}}$, $G_j = 2\pi q_j p_j LM_s/\gamma$, where $\gamma$ is the gyromagnetic ratio. The topological charges $q_1 = q_2 = 1$ are considered below, which describes the single vortices observed experimentally in soft magnetic cylinders at remanence.[1-3] The gyrovector, an intrinsic property of the vortex,[17] is of principal importance in describing the vortex dynamics and can be calculated by integration over the vortex core.[5] The second terms in Eq. (1) describe the restoring forces acting on vortices that are shifted from their equilibrium positions at the dot centers. The restoring force is a consequence of the finite lateral size of the dot and is directed toward the dot center. For sub-micron dot radii, the dot magnetostatic energy (shifting the vortex induces magnetic charges) provides the main contribution to the shifted-vortex energy $W(\mathbf{X}_1, \mathbf{X}_2)$.

The static energy dependence of a vortex on its core position was calculated for a cylindrical dot [3,5] on the basis of a model with no magnetic side surface charges,[19] where the complex function $w(\zeta,\bar\zeta) = \tan(\Theta(x,y)/2)\exp(i\Phi(x,y))$, with the complex variable $\zeta = (x+iy)/R$, was used. The function $w(\zeta,\bar\zeta)$ has the form[6] $w(\zeta,\bar\zeta) = f(\zeta)$ if $|f(\zeta)|<1$ (within the vortex core) and $w(\zeta,\bar\zeta) = f(\zeta)/|f(\zeta)|$ if $|f(\zeta)| \geq 1$, where $f(\zeta)$ is an appropriate analytical function. In this case the function $f(\zeta) = (1/c)(i\zeta C + (a - \bar a \zeta^2)/2)$ was used, where $c=b/R$ is the relative core radius, which corresponds to the "side surface charge free" model $[(\mathbf{m}\cdot\mathbf{n})_S = 0]$.[19] The function $f(\zeta)$ for the complex parameter $a$ ($|a|<1$) describes magnetization field of two vortices. The first vortex has its center at $\zeta_1 = iaC/2$, and a second image vortex is centered at $\zeta_2 = 2iaC/|a|^2$. Their coordinates are connected by the inversion transformation $\zeta_1 \bar\zeta_2 = 1$. It is convenient to consider the complex parameter $a = a' + ia''$ as a 2D vector $\mathbf{a} = (a_x, a_y)$ with components $a_x = a'$ and $a_y = a''$. The vector $\mathbf{a}$ corresponds to the volume-averaged magnetization of the shifted vortex $\langle \mathbf{m}(\mathbf{r}) \rangle_V = \mathbf{a}/3$. It is perpendicular to the vortex core shift vector $\mathbf{s}=\mathbf{X}/R$ and can be expressed as $\mathbf{a} = -2C\,\mathbf{z}\times\mathbf{s}$. The value $a=0$ corresponds to a centered vortex.[3]

For a small displacement of the vortex center from its equilibrium position ($\mathbf{X}=0$), one can write $W_j(\mathbf{X}_j) = W_j(0) + 1/2\kappa X_j^2$ for each $F$ layer, where the stiffness coefficient $\kappa$ can be determined from the decomposition of the vortex energy in a power series of $a$.[5] For the present model, the interlayer magnetostatic coupling can be represented as the sum of contributions from the volume and face surface magnetic charges. Due to the small size of the vortex core $c\ll 1$ (where the face surface charges are non-zero), the contribution of the face surface charges has been neglected. The interaction energy of the volume charges in the first and second dots can be calculated explicitly in the symmetric form:

$$W_{int}(\beta,d) = 2\pi^2 M_s^2 R^3 F(\beta,d)(\mathbf{a}_1 \cdot \mathbf{a}_2), \qquad (2)$$

where $F(\beta, d_s) = \int_0^\infty dt\, t^{-2} \exp(-td_s)(1-\exp(-\beta t))^2 I^2(t)$, $I(t) = \int_0^1 dx\, xJ_1(tx)$, $\beta = L/R$, $d_s = d/R$.

Using the relations $\mathbf{a}_j = -2C_j\, \mathbf{z} \times \mathbf{s}_j$, the total magnetostatic energy density for a trilayer F/N/F disk with shifted vortices centered at $\mathbf{X}_1$ and $\mathbf{X}_2$, and chiralities $C_1$ and $C_2$ can be expressed as

$$w(\mathbf{X}_1, \mathbf{X}_2) = \frac{1}{2}\kappa_1 X_1^2 + \frac{1}{2}\kappa_2 X_2^2 + \mu(\mathbf{X}_1 \cdot \mathbf{X}_2), \quad \text{where} \quad \mu = 8\pi^2 M_s^2 RF(\beta, d_s) C_1 C_2. \tag{3}$$

For simplicity, we assume F disks with the same parameters ($L$, $\gamma$ and $M_s$), but this approach can be readily generalized for the case of layers with different parameters. The coefficients $\kappa_1 = \kappa_2 = \kappa$ were calculated[5] as $\kappa = 4\pi LM_s^2 \left[4\pi F_v(\beta) - 0.5(R_0/R)^2\right]$, where $F_v(\beta) = \int_0^\infty dt\, t^{-1} f(\beta t) I^2(t)$ is a function of the dot aspect ratio $\beta$, $f(x) = 1 - (1-\exp(-x))/x$, and $R_0 = (2A/M_s^2)^{1/2}$ is the exchange length.

The cases of equal ($p_1 p_2 = +1$) or opposite ($p_1 p_2 = -1$) vortex core polarizations are considered separately. The case $p_1 p_2 = +1$ corresponds to the lowest magnetostatic energy and will be considered below. For this case the equations of motion can be diagonalized using the transformation $\mathbf{X} = (\mathbf{X}_1 + \mathbf{X}_2)/2$, $\boldsymbol{\xi} = \mathbf{X}_1 - \mathbf{X}_2$. The equation of motion for the new collective coordinates $\mathbf{X}(t)$, $\boldsymbol{\xi}(t)$ are reduced to the equations of a harmonic oscillator with the eigenfrequencies

$$\omega_X = \frac{1}{G}(\kappa + \mu), \quad \omega_\xi = \frac{1}{G}(\kappa - \mu). \tag{4}$$

Note that $\omega_X > \omega_\xi$ for the same vortex chiralities $C_1 C_2 = +1$ and $\omega_X < \omega_\xi$ for opposite chiralities $C_1 C_2 = -1$. Only the case of $C_1 C_2 = -1$ is considered here because this is the state that is observed in micromagnetic simulations when the system is allowed to relax from a saturated state. The total volume-averaged magnetization $\langle \mathbf{m} \rangle_V = -\hat{\mathbf{z}} \times (C_1 \mathbf{X}_1 + C_2 \mathbf{X}_2)/3R$ always shows oscillations at the highest frequency $\max(\omega_X, \omega_\xi)$. The low-frequency mode can only be observed if the average magnetization of the individual layers is considered. The vortex eigenfrequency for well-separated layers ($d \to \infty$) $\omega_0 = \kappa/G$ corresponds to the eigenfrequency of a dot with thickness $L$ because $\mu(d = \infty) = 0$. As the spacer thickness approaches $d=0$, the highest frequency $\max(\omega_X, \omega_\xi)$ approaches the eigenfrequency of an isolated dot with thickness $2L$, whereas the lowest frequency is very small. It can be shown that $\min(\omega_X, \omega_\xi) = 0 + O(\beta^2)$ at $\beta \ll 1$.

For typical dot sizes with $M_s = 800$ G (FeNi), the eigenfrequency $\omega_0/(2\pi)$ of an isolated dot lies in the GHz range.[5] The calculated eigenfrequencies of F/N/F dot ($p_1 p_2 = +1$) can be written as:

$$\omega_X = \omega_M \left[ 2F_v(\beta) + \frac{C_1 C_2}{\beta} F(\beta, d) \right], \quad \omega_\xi = \omega_M \left[ 2F_v(\beta) - \frac{C_1 C_2}{\beta} F(\beta, d) \right], \quad \omega_M = 4\pi\gamma M_s. \quad (5)$$

The dependences of the vortex eigenfrequencies on the F/N/F geometrical parameters $d$ and $\beta$ are shown in Fig. 2. The considerable splitting of the vortex eigenfrequencies (Fig. 2a) is the result of the interlayer dynamic magnetostatic interactions. The vortex chiralities determine the symmetry of the eigenmodes of oscillations of the layer magnetizations. The vortex core trajectories are a superposition of a fast circular motion around the dot center and a slow change of the circular orbit position (Fig. 3).

To further analyze the vortex dynamics we used micromagnetic simulations based on the Landau-Lifshitz Gilbert equation.[20] Diameters $D$ of 500, 750, and 1000 nm were used with a spacer thickness $d$ of 1 nm and a F layer thickness $L$ of 40 nm. The small values of $d$ ensures strong magnetostatic coupling between the disks. $L = 40$ nm is thick enough to favor the vortex state but thin enough to satisfy the assumption of magnetization uniformity along thickness. Magnetic constants appropriate for bulk Permalloy ($Ni_{80}Fe_{20}$) were used: exchange constant $A = 1.05$ μerg/cm, $M_s = 800$ emu/cm$^3$, and $\gamma = 18.47$ MHz/Oe. Magnetic anisotropy was neglected and no interlayer exchange was incorporated into the model. The models were defined using cells 3-4 nm in size with thickness $L$.

Firstly, the models were allowed to relax from an in-plane saturated state with a large damping parameter ($\alpha = 0.8$) in order to determine the remanent state. For diameters of 500 and 1000 nm the disks relaxed into the double vortex structure illustrated in Fig. 1. The two F layers support centered vortices of opposite chirality and identical polarization. Next, a static in-plane magnetic field of 100 Oe was applied in the $Ox$ direction. The double vortex was then allowed to relax into a new equilibrium state ($\alpha = 0.8$), that is characterized by equal displacement of the vortices in opposite directions along the y-axis. Finally, the dynamic response of the system was simulated in response to an in-plane field profile: 100 Oe in the $x$-direction held constant for 500 ps and then reduced linearly to zero over 100 ps. After removing magnetic field, the vortex cores oscillate back to their remanent states in the centers of the disks. For the limiting case of zero damping ($\alpha=0$), the centers of the vortices will continue to oscillate along trajectories that are a combination of two circular motions with the eigenfrequencies given by Eq. (5). For the dynamic simulations a small damping constant was used ($\alpha = 0.008$ for realistic simulations, $\alpha = 0.001$ for improved frequency estimates). The directions of rotation of the vortex cores are the same, as defined by the signs of the core polarizations $p_1$, $p_2$ in each layer.

The frequencies of the vortex translation modes determined using the simulations are shown in Fig. 2b (open symbols). The analytical results based on the "two-vortices" model agree reasonably well with the micromagnetic LLG calculations for all the values of $\beta$. The results of selected dynamic simulations are shown in Fig. 4. The total volume-averaged magnetization for a 500 nm-diameter disk (Fig. 4a) shows damped oscillations as a function of time with a peak frequency of 1.0 GHz, which matches the frequency of the vortex core translational mode for a simulated, single 80 nm-thick Py disk. Examination of the volume-averaged response of the individual magnetic layers (Fig. 4b), however, reveals an additional low-frequency mode (~75 MHz) in agreement with Eq. (5). Tracking the motion of the vortex cores in both layers provides further insight into the origin of these two modes. Figures 4c and 4d show that the difference in vortex core positions $\xi = X_1 - X_2$ (Fig. 4c) corresponds to a high frequency translational motion, while the mean position of the cores $X = (X_1 + X_2)/2$ (Fig. 4d) oscillates about the disk center at a much lower frequency, in full agreement with the analytical predictions. This lower frequency mode originates from dynamic interaction between vortices and represents small amplitude oscillations of the mean core position.

In summary, the dynamic properties of magnetically soft submicron ferromagnetic tri-layer *F/N/F* dots of the opposite chirality vortex state have been explored via analytical and numerical calculations. Calculations using the "side charges free" model reveal two eigenfrequencies for dots with the same core polarizations corresponding to coupled translation modes of the two vortices. Thus, we predict the existence of two excitation modes with a frequency difference. The vortex center trajectories can be understood in terms of a high frequency (~1 GHz) circular (no damping) or spiral motions (with damping), and a much lower frequency rotation of the mean core position superimposed. The analytical description is in good qualitative and reasonable quantitative agreement with micromagnetic simulations. The eigenfrequencies of the vortex translation modes depend on the dot aspect ratio $\beta$ and spacer thickness *d* and should be observable by modern experimental techniques.

This work was supported by the U.S. Department of Energy, BES Material Sciences under Contract No. W-31-109-ENG-38. K. B. thanks the NSERC of Canada for a fellowship.

**Figure captions**

Fig. 1. Diagram showing the F/N/F tri-layer disk structure.

Fig. 2.

The vortex eigenfrequencies vs. a) the interdot separation ($\beta = 0.1$), and b) the dot aspect-ratio. The solid and dash-dotted lines show the eigenfrequencies ($\omega_X, \omega_\xi$) for $p_1 p_2 = +1$ as calculated from Eq. (5). In a), the dotted line corresponds to the eigenfrequency of an isolated dot. The interdot separation is $d=$ 1 nm, and the dot thickness is $L=$ 40 nm. The micromagnetic data for $2R =$ 500, 750, and 1000 nm are shown as symbols.

Fig. 3.

The trajectory of a vortex core precessing around the remanent equilibrium position for $p_1 p_2 = +1$ and $C_1 C_2 = -1$ is shown over a) 16 ns and b) 48 ns (corresponding to the period of the low-frequency eigenmode) for a $F/N/F$ dot geometry of 40/1/40 nm, $R=$ 500 nm. The amplitudes and frequencies of the coordinates $\mathbf{X}$ (6 nm and 21 MHz) and $\xi$ (76 nm and 0.51 GHz) oscillations were taken from the micromagnetic simulations.

Fig. 4.

Simulation results for $p_1 p_2 = +1$ and $C_1 C_2 = -1$. Volume-averaged magnetization (y-component) of Py/N/Py disk with $L = 40$ nm, $d = 1$ nm, $2R = 500$ nm, $\alpha = 0.008$ for a) the full disk, and b) the individual magnetic layers. c) The difference between core positions in the two layers shows high frequency oscillations, while d) the mean core position varies at a much lower frequency (core positions are shown for $2R = 1000$ nm, $\alpha = 0.001$).

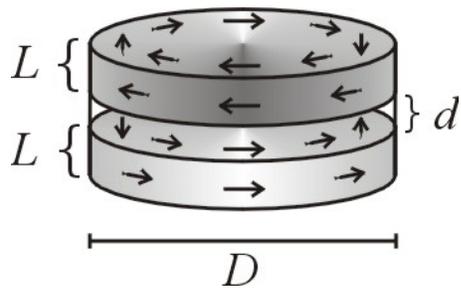

Fig. 1 to the manuscript by K.Yu. Guslienko et al. "Dynamics of coupled vortices…"

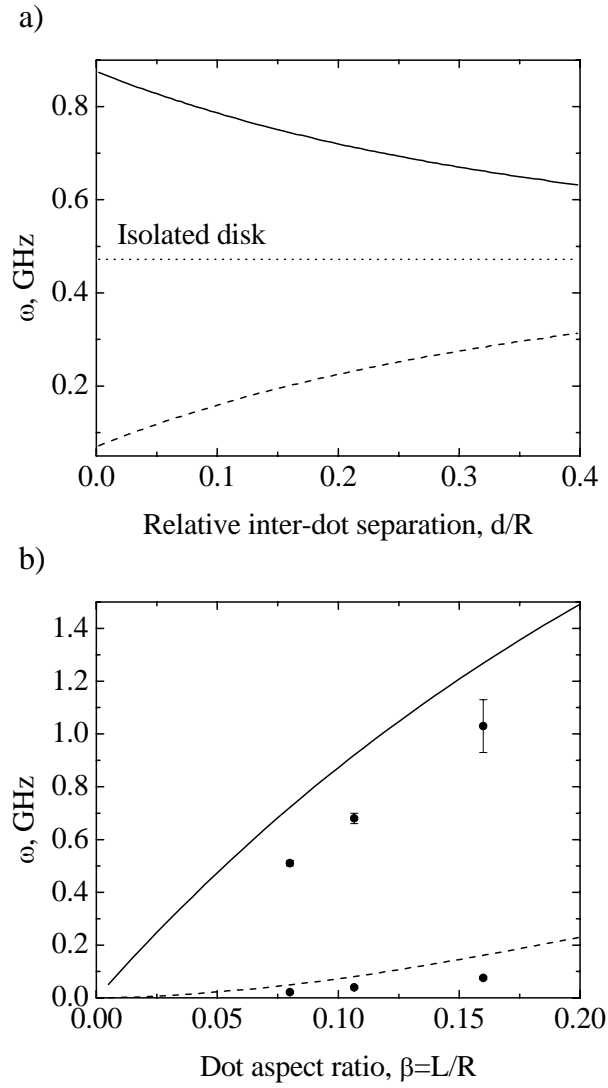

Fig. 2 to the manuscript by K.Yu. Guslienko et al. "Dynamics of coupled vortices…"

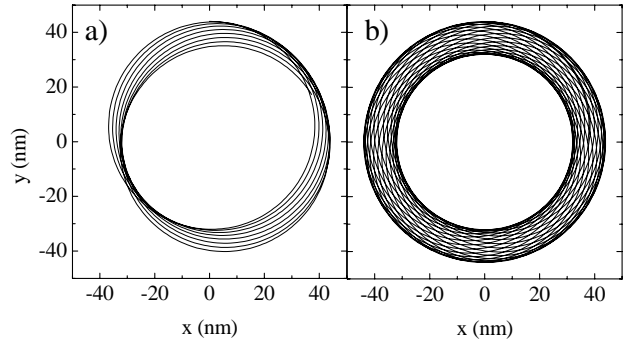

Fig. 3 to the manuscript by K.Yu. Guslienko et al. "Dynamics of coupled vortices…"

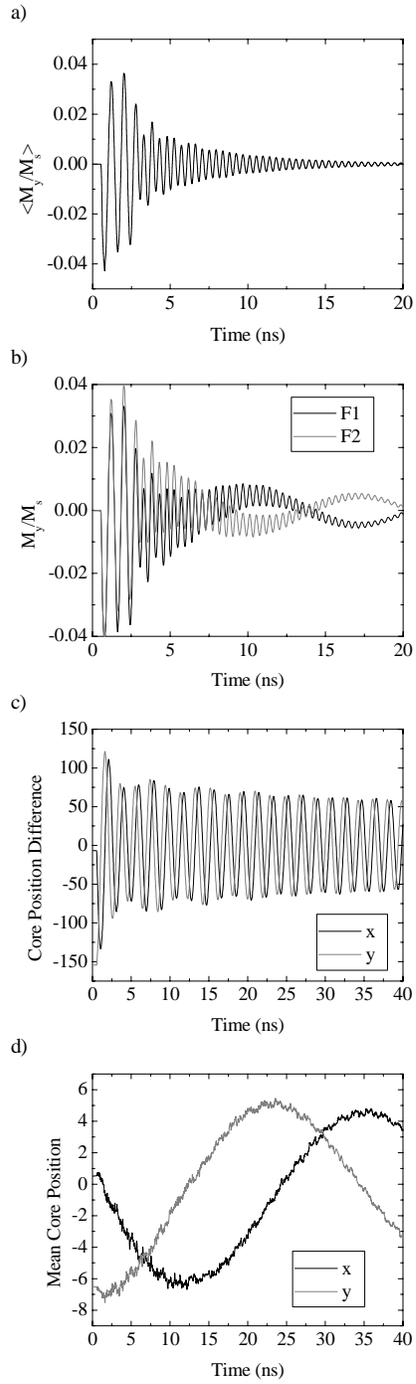

Fig. 4 to the manuscript by K.Yu. Guslienko et al. "Dynamics of coupled vortices…"